\documentclass[12pt]{article}
\usepackage{amssymb}

\newcommand{\be}{\begin{equation}}
\newcommand{\ee}{\end{equation}}
\newcommand{\bea}{\begin{eqnarray}}
\newcommand{\eea}{\end{eqnarray}}
\newcommand{\<}{\langle}
\renewcommand{\>}{\rangle}
\def\bsigma{\mbox{\protect\boldmath $\sigma$}}

\def\hatp{\hat p}

\newcommand{\beq}{\begin{eqnarray}}
\newcommand{\eeq}{\end{eqnarray}}

\newcommand{\R}{\hbox{{\rm I}\kern-.2em\hbox{\rm R}}}
\newcommand{\reff}[1]{(\ref{#1})}
\def\ltapprox{\lesssim}
\def\gtapprox{\gtrsim}

\newcommand{\Z}{\mathbb{Z}}

\begin{document}

\title{Asymptotically free models and discrete non-Abelian groups}

\author{
  { Sergio Caracciolo }              \\
  {\small\it Scuola Normale Superiore, INFM and INFN -- Sezione di
  Pisa}  \\[-0.2cm] 
  {\small\it I-56100 Pisa, ITALIA}          \\[-0.2cm]
  {\small Internet: {\tt Sergio.Caracciolo@sns.it}  }
  \\[-0.1cm]  \and
  { Andrea Montanari}              \\
  {\small\it Laboratoire de Physique Th\'{e}orique de l'Ecole Normale
  Sup\'{e}rieure\footnote {UMR 8549, Unit{\'e}   Mixte de Recherche du 
Centre National de la Recherche Scientifique et de 
l' Ecole Normale Sup{\'e}rieure. }  }
  \\[-0.2cm]
  {\small\it 24, rue Lhomond, 75231 Paris CEDEX 05, FRANCE}        \\[-0.2cm]
  {\small Internet: {\tt Andrea.Montanari@lpt.ens.fr}}
  \\[-0.1cm]  \and
  { Andrea Pelissetto}              \\
  {\small\it Dipartimento di Fisica and INFN -- Sezione di Roma I}
  \\[-0.2cm]
  {\small\it Universit\`a degli Studi di Roma ``La Sapienza"}        \\[-0.2cm]
  {\small\it I-00185 Roma, ITALIA}          \\[-0.2cm]
  {\small Internet: {\tt Andrea.Pelissetto@roma1.infn.it}}   \\[-0.2cm]
  {\protect\makebox[5in]{\quad}}  % To force authors' names to be written
                                  %   vertically, one above another.
                                  % (\author seems to put them side-by-side
                                  %   if there is room.)
  \\
}
%\vspace{0.5cm}

%%\date{}   % REMEMBER TO PUT DATE IN FINAL VERSION!!!

\maketitle
\thispagestyle{empty}   % Suppress page number on front page.

\abstract{We study the two-dimensional renormalization-group
  flow induced by perturbations that reduce the
  global symmetry of the $O(3)$ $\sigma$-model to the discrete symmetries 
  of Platonic solids. We estimate the value of the correlation
  length at which 
  differences in the behaviour of the various models should be 
  expected. For the icosahedron model with nearest-neighbor interactions, 
  we find $\xi\gtapprox 200$.
  We provide an explanation for the recent numerical results of 
  Patrascioiu and Seiler and of Hasenfratz and Niedermayer.}

\clearpage

\section{Introduction}

Quantum field theories with a non-Abelian continuous symmetry group of
invariance play a major role both in particle and in condensed-matter
physics. Two cases are of particular interest: two-dimensional 
spin models with non-Abelian global symmetry group, 
and four-dimensional gauge theories with local non-Abelian gauge invariance.
According to the common wisdom these two cases share the peculiar
feature of asymptotic freedom (AF).

In the lattice formulation it is quite easy to replace the continuous
group by one of its discrete subgroups.  In this case, due to the
discreteness of the group, the action has a finite gap and  at
least a freezing transition is expected.\footnote{In some models, 
for instance in some $Z(N)$ models,
between the ordered and the disordered phase, there may also be 
an intermediate massless phase~\cite{Elitzur}.} Nonetheless, at
large enough temperature one expects only small differences and indeed, 
finite subgroups have been used in Monte
Carlo updates to simulate continuous groups~\cite{Rebbi,Bhanot}.

When the symmetry group is Abelian it may happen that the discrete
symmetry is enlarged to a continuous one. The massless intermediate 
phase of $Z(N)$ models with $N>4$ is the same of the
$O(2)$ model for low enough temperature~\cite{K,J}.

It has also been suggested that a similar phenomenon occurs in
non-Abelian models. Patrascioiu and
Seiler~\cite{Patrascioiu:1998ds,Patrascioiu:1998ww,Patrascioiu:2000mw,
  Patrascioiu:2000ek} 
have often criticized the conventional wisdom on AF for the continuous
group and proposed an alternative scenario in which a low-temperature
massless phase appears. If this possibility really happens it is plausible to
accept the idea that, for example, in $d=2$ the $O(3)$ model is in the same
universality class of the dodecahedron spin model. 

In order to test this conjecture several large-scale simulations 
have been performed ~\cite{Hasenfratz1,Hasenfratz2,Patrascioiu:2000mw}.
In particular, the finite-size scaling curve
for the second-moment correlation length measured
in~\cite{Caracciolo:1995ud,Caracciolo:1995ed}  and the
renormalized coupling for the $O(3)$ model have been compared with the 
results obtained for discrete spin models with different discrete subgroups
and nearest-neighbor interactions.
It was found that the icosahedron model and the $O(3)$ model
 are practically indistinguishable
at present-day values of $\xi$. This is not totally surprising: 
after all, for $\beta$ small enough one expects only tiny differences
since in the presence of large fluctuations the discreteness of the 
spin space should not play an important role (this was indeed the motivation 
of Refs. \cite{Rebbi,Bhanot}). However, what is more surprising 
is that the discrepancy seems to decrease as $\beta$ increases (see the 
results of Ref. \cite{Hasenfratz1} for the renormalized zero-momentum
four-point coupling), 
while naively one would have expected the opposite.

These numerical results have been interpreted as evidence that the 
$O(3)$ and the icosahedron model have the same continuum limit.
In this paper, we will show that this conclusion is in contrast
with the common theoretical understanding of the $O(3)$ model: If the 
continuum limit of the $O(3)$
$\sigma$ model is correctly described by the perturbative 
renormalization group (RG), then the icosahedron and the $O(3)$ model belong 
to {\em different} universality classes, contrary to what suggested in 
Refs.~\cite{Hasenfratz1,Hasenfratz2}.

For this purpose,
we study the effect of  perturbations that break the
$O(3)$ symmetry down to a discrete subgroup and show that
any such perturbation is a relevant perturbation
that modifies the universal behavior.
A similar analysis for the cubic symmetry was performed by Pelcovits
and Nelson~\cite{PN} and in the context of the $XY$ model in Ref. \cite{K}.
We then try to provide an explanation to the numerical data.     
We analyze a model with icosahedral symmetry that interpolates between
the nearest-neighbor $O(3)$ model ($h_6=0$) and the standard 
icosahedron model ($h_6=+\infty$), using perturbation theory.
When $h_6$ is small, we find for $\xi\ltapprox 200$ a behavior that is 
similar to that observed numerically: the difference between the 
icosahedral and the $O(3)$ model decreases as $\beta$ increases.
However, this apparent convergence is misleading, because for $\xi
\gtapprox 200$ the difference 
between the two models {\em increases}\/ as $\beta$ increases,
with the result that the two models {\em do not}\/ lie in the same
universality class.
On the basis of these results we predict that the difference between 
the standard icosahedron model considered 
in~\cite{Hasenfratz1,Hasenfratz2,Patrascioiu:2000mw} and the $O(3)$ 
model should probably become visible only for $\xi\gtapprox200$. 

We stress that all our arguments assume the validity of the 
perturbative RG for the $O(3)$ model.
If this is not correct, it is possible that the 
icosahedron and the $O(3)$ models have the same universal behavior as 
conjectured by Patrascioiu and Seiler.

\section{Renormalization-group analysis of perturbations with discrete 
symmetry}

In this Section we want to perform a RG analysis 
of the discrete models.  
For this purpose we consider the lattice Hamiltonian
\be
\beta H^{\rm latt}  =  {\beta\over 2} \sum_{x\mu} (\Delta_\mu\bsigma)^2 - 
            h_n \sum_x  I_n(\bsigma_x),
\label{Hmista}
\ee                                                                   
where $\bsigma_x$ is a unit vector in $\R^3$, 
\be
\Delta_\mu \bsigma_x = \bsigma_{x+\mu} - \bsigma_x ,
\ee
and $\mu=1,2$ are the positive directions on a square lattice with
lattice spacing $a=1$. Here, $I_n(\bsigma_x)$ is a polynomial of 
$\bsigma_x$ that is invariant under the action of the discrete group 
and belongs to an irreducible representation of the $O(3)$ group, i.e. 
it has a well-defined $O(3)$ spin  $n$. Such polynomials can be found for all 
discrete subgroups that correspond to Platonic solids. 
The lowest-rank tensors for each subgroup are 
obtained in App. \ref{inv}. For the tetrahedron, 
the cube (or the octahedron), and for the icosahedron (or the dodecahedron),
the lowest-rank invariant tensors have spin $n=3$, $4$, and 6. 
Explicitly, we have for the three cases respectively
\begin{eqnarray}
I_3 & = & -i\,3\, \sqrt{2 \pi\over 35} \left( Y_{3,2} - Y_{3,-2}\right),  
\nonumber \\
I_4 & = & -\sqrt{\pi}\,\left[ Y_{4,0} +  \sqrt{5\over 14} \left(
    Y_{4,4} + Y_{4,-4} \right) \right],
\nonumber \\
I_6 & = &  \sqrt{4 \pi\over 13}\,\left[ Y_{6,0} -  \sqrt{7\over 11} \left(
    Y_{6,5} - Y_{6,-5}\right ) \right]  .
\label{def-Ipoly}
\end{eqnarray}
Here, we have introduced the spherical harmonics
\be
Y_{l,m}(\theta,\phi) = \sqrt{ {2 l+1 \over 4 \pi} {(l-m)!\over (l+m)!}
  }\, P_{l,m}(\cos \theta )\, e^{i m \phi} ,
\ee
where $ P_{l,m}(x)$ are the associated Legendre functions, and  we have 
written the three-dimensional spin $\bsigma$ in terms of polar coordinates
$\theta$ and $\phi$. 
These tensors are not unique, since the normalization is arbitrary and they
are defined modulo a rotation.\footnote{In Eq. \reff{def-Ipoly} we
  have defined  
$I_n$ so that each polynomial has maximal value one on the vertices of
the tetrahedron, 
cube, and icosahedron respectively. The minimal value 
(maximal in absolute value) 
of $I_4$ on the vertices of the 
octahedron is $-3/2$ and of $I_6$ on the vertices of the 
dodecahedron is $-5/9$. Of course, the normalization of $I_n$ 
is irrelevant for our discussion.}
For the purpose of showing that discrete-subgroup perturbations are relevant, 
the value of $n$ plays no role and one can use any invariant tensor. 
One could also consider tensors that do not have definite spin and are
sums of invariant $O(3)$-irreducible tensors. This would only make 
the discussion more cumbersome without changing the physical results.

The partition function is defined as usual by
\be
Z \, =\, \int  e^{-\beta H^{latt}}  \prod_x \hbox{\rm
d}^N\bsigma_x \, \delta(\bsigma_x^2 -1) \;.
\ee
The Hamiltonian \reff{Hmista} interpolates between the $O(3)$ model 
($h_n=0$) and the discrete-symmetry model ($|h_n|=\infty$).\footnote{
Note that, if a discrete model is obtained fo $h_n = +\infty$, 
the dual model (see App. \ref{App1.1} for the definition of duality) 
is obtained for $h_n = - \infty$.}

If, as claimed in Refs.~\cite{Hasenfratz1,Hasenfratz2}, 
the discrete-symmetry model\footnote{Strictly speaking, 
the claim has been made only for $|h_n|= \infty$. However, it is difficult 
to imagine a scenario in which the theory is not $O(3)$ invariant 
for $h_n$ small and recovers the continuous invariance by increasing the 
strength of the discrete-symmetry term.} is also AF,  in the RG
language, this means that the added term is an irrelevant perturbation. 
Let us perform a standard RG
calculation around the theory with $h_n=0$. If
$G^{(p)}(k_1,\ldots,k_p;\beta,h_n)$ is  
the connected $p$-point 
correlation function, we can perform an expansion in powers of 
$h_n$, i.e. rewrite
\be
G^{(p)}(k_1,\ldots,k_p;\beta,h_n) = \, 
  \sum_{q=0} {1\over q!} h_n^q G^{(p,q)}(k_1,\ldots,k_p;0,\ldots,0;\beta,0),
\ee
where $G^{(p,q)}(k_1,\ldots,k_p;l_1,\ldots,l_q;\beta,0)$ is the correlation 
function with 
$p$ fields and $q$ insertions of the breaking operator at momenta $l_1$, 
$\ldots$, $l_q$, 
computed for $h_n=0$, i.e. in the $O(3)$-model. Such correlation functions 
cannot be computed directly in perturbation theory
for $l_1=\ldots=l_q=0$,
since the momenta of the insertions are zero and perturbative 
expansions\footnote{Even for large momenta, perturbative expansions
are well-defined only for $O(3)$-invariant
quantities. Thus, in order to have an infrared-finite perturbation theory
one must rewrite each correlation function in terms of $O(3)$-invariant
expressions.
For instance, to compute correlations of $I_n$ we can use the
identity $\< Y_{l_1,m_1} (\bsigma_0) Y_{l_2,m_2} (\bsigma_x)\> =
\delta_{l_1,l_2} \delta_{m_1,m_2} \< P_{l_1} (\bsigma_0\cdot \bsigma_x)\>/
(4\pi)$, where $P_l(x)$ is a Legendre polynomial. The right-hand side
is infrared-finite.}
are valid only for large momenta, i.e. for distances much 
smaller than the correlation length.
Indeed, we are expanding around an ordered configuration and this is correct
only for $|x|\ll \xi$ because of the Mermin-Wagner theorem. However,
within the usual theoretical framework, there is a standard way out. 
Consider $G^{(p,q)}(k_1,\ldots,k_p;l_1,\ldots,l_q;\beta,0)$ in the perturbative
regime (large momenta) and use perturbation theory to derive the 
RG equation
\be
\left[ -a {\partial\over \partial a} + W(t) {\partial\over \partial t} + 
q \gamma^{(n)}(t) + {p\over2} \gamma(t)\right] 
G^{(p,q)}(k_1,\ldots,k_p;l_1,\ldots,l_q;\beta,0) = 0,
\ee
where $t\equiv1/\beta$, 
$W(t)$, $\gamma^{(n)}(t)$, and $\gamma(t)$ are respectively
the lattice $\beta$-function and the lattice anomalous dimensions of 
$I_n$ and of the field. The crucial assumption is that this equation 
is valid for all momenta, even outside the strictly perturbative 
region.\footnote{
The correlation functions are computable in perturbation theory 
if we introduce an infrared cutoff, for instance if we work in 
a finite volume. However, they will be infrared divergent, so that one cannot 
take the infinite-volume limit naively. In the standard theoretical framework,
one uses the perturbative expressions to derive an RG equations
and then assumes the such equations are also satisfied by the 
infinite-volume quantities. Whatever the procedure, the result 
is identical.} 

Taking into account the scaling dimension of the 
correlation function, at zero external momenta ($k_i=l_i=0)$, we have 
\be
G^{(p,q)}(0;0;\beta,0) = A^{(p,q)} G^{(p,0)}(0;\beta,0) 
  \exp\left[q \int^t_{t_0} {2 - \gamma^{(n)}(s)\over W(s)}\,
    ds\right],
\ee
where $A^{(p,q)}$ is a non-perturbative constant. Therefore, 
we obtain finally
\be
G^{(p)}(0;\beta,h_n) = 
  G^{(p)}(0;\beta,0) \sum_{q=0} {1\over q!} A^{(p,q)} h^q_n
   \exp\left[q \int^t_{t_0} {2 - \gamma^{(n)}(s)\over W(s)}\, ds\right] .
\label{Gn-funzioneh}
\ee
By using the perturbative RG, we have been able to factor out 
the $h$ dependence of the correlation function. Now, for $t\to 0$ 
we find immediately 
\be
G^{(p)}(0;\beta,h_n) \approx 
  G^{(p)}(0;\beta,0) \sum_{q=0} {1\over q!} A^{(p,q)} \left(h_n t^{\rho_n}
   \exp\left[{4\pi \over t}\right]\right)^q,\label{risultato}
\ee                                                                            
where $\rho_n$ is an easily computable exponent.
The correction term diverges for $t\to 0$, showing that the breaking term 
is a relevant interaction in the RG sense. 

Equivalently, one can imagine of considering a scale-dependent 
renormalized coupling $h_n(s)$. Then, the RG flow has the form
\be
{1\over h_n(s)} {d h_n(s)\over ds} = 2 -
\gamma^{(n)}\left[t(s)\right] ,
\ee
where $\exp(-s)$ is the change of the scale and $t(s)$ is the running 
coupling constant. 
Since $\gamma^{(n)}(t)$ vanishes for $t\to 0$,
any perturbation of this type is relevant in the continuum limit.
We would like to point out that this is not unexpected. Since in two
dimensions  
the field $\bsigma$ is dimensionless, any polynomial in $\bsigma$ is 
a relevant operator.

Equations \reff{Gn-funzioneh} and \reff{risultato} deserve some 
additional comments. First of all, they give an expansion of 
$G^{(p)}(0;\beta,h_n)$ in powers of the scaling variable
\be 
z \equiv h_n t^{\rho_n} \exp(4\pi/t), 
\label{def-z}
\ee
and are therefore valid only for $z\ll 1$. However, for our purposes, 
the only relevant information we obtain from Eq. \reff{risultato} 
is that the correct scaling variable is $z$, i.e., that we can define 
the limit $t\to 0$, $h_n\to 0$ at fixed $z$, obtaining 
\be
G^{(p)}(0;\beta,h_n) = G^{(p)}(0;\beta,0) \Phi^{(p)}(z),
\label{crossover}
\ee
where $\Phi^{(p)}(z)$ is a nonperturbative crossover function. 
Eq. \reff{crossover} can be derived directly from the 
RG equation 
\be
\left[ -a {\partial\over \partial a} + W(t) {\partial\over \partial t} + 
\gamma^{(n)}(t) h_n {\partial\over \partial h_n}  + {p\over2} \gamma(t)\right] 
G^{(p)}(k_1,\ldots,k_n;\beta,h_n) = 0,
\label{RGeq-generale}
\ee
which we assume to be valid for all momenta in a neighborhood of 
$h_n=t=0$. Of course, we can only compute the scaling behavior of the 
correlation functions, but not their explicit expressions. 
Our ignorance of the long-wavelength physics 
is encoded here in the nonperturbative nature of the function $\Phi^{(p)}$. 

Our result~\reff{crossover} provides what in statistical mechanics is called a 
crossover scaling function for a fixed point perturbed 
by two relevant interactions (or, in this case, by one relevant interaction
$h_n$ and one marginally relevant interaction $t$):  namely, it gives
the leading behavior in the limit $t \to 0$, $h_n \to 0$ with $z$
fixed. Note that Eq. \reff{crossover} can also be used to predict 
the behavior of the phase-transition line for $h_n$ small 
\cite{KNF-76,Amit}, if such a 
transition exists.\footnote{Eq. \reff{crossover} is valid for any
perturbation and  any $n$, even if it is not associated to a discrete 
group. For instance, one can take $n=1$ and $I_1(\bsigma) = \cos\theta$,
obtaining the usual magnetic perturbation. In this case, 
Eq. \reff{crossover} is the usual crossover equation (for $p=1$ it is 
usually called equation of state) reported in 
textbooks, see, e.g., Ref. \cite{ZinnJustin}. 
Instead, Eq. \reff{phase-transition} that will be derived below makes no sense 
since at fixed $h_1$ there is no phase transition.}  Indeed,
if $\beta_c(h_n)$ is the critical point of the theory at fixed $h_n$, 
for $h_n$ small we should have, for $h_n$ small, 
\be
h_n = z^* \beta_c(h_n)^{\rho_n} \exp[-4 \pi \beta_c(h_n)],
\label{phase-transition}
\ee
where $z^*$ is a nonperturbative constant.

% Finally, let us discuss the nature of the renormalized continuum theories
% that can be obtained by taking the limit $h_n\to 0$, $t\to 0$ at $z$ fixed. 
% For $z=0$  we obtain the usual $O(3)$ renormalized massive theory,  
% for $z=z^*$ we obtain the massless discrete-group fixed point theory, 
% while for $0<z<z^*$ we obtain intermediate massive renormalized theories.

We want now to explain the numerical results of Refs.
~\cite{Hasenfratz1,Hasenfratz2,Patrascioiu:2000ek}, who found 
that the difference in behavior between the standard $O(3)$ model
and the discrete model was decreasing as $\beta$ increased. A possible 
explanation of this phenomenon is 
that the RG flux first reduces the size of the 
perturbation which then increases again as $\beta$ increases. 
Since in the high-temperature regime, one expects indeed the two models
to be quite similar, this could explain the fact that they are 
numerically indistinguishable at the values of $\beta$ that can 
be simulated today. 

To make this picture more quantitative, let us consider 
the Hamiltonian \reff{Hmista} with $h_n$ and $z$ small, so that we can use 
Eq. \reff{Gn-funzioneh}. Then, suppose 
that there exists $t^{\rm eff} = 1/\beta^{\rm eff}$ 
such that $\gamma^{(n)}(t^{\rm eff}) = 2$ and 
$2 - \gamma^{(n)}(t^{\rm eff}) < 0$ for $\beta < \beta^{\rm eff}$. 
In this case,
Eq. \reff{Gn-funzioneh} would predict the following behavior.
For $\beta$ small, the difference would apparently decrease as $\beta$ 
increases, which could seem to indicate that the interaction is irrelevant.
However, as soon as $\beta$ becomes larger than $\beta^{\rm eff}$ 
the discrepancy starts increasing again. 
Now, notice that the discrete model, in the vicinity of 
the $O(3)$ fixed point, will generate 
perturbations of arbitrary spin. However, $\gamma^{(n)}$ increases
with $n$, and thus the most relevant perturbation is associated to 
$I_n$ with the smallest possible value of $n$. Therefore, we should 
consider $n=3$, 4, 6 for the tetrahedron, the cube and the octahedron, 
the dodecahedron and the icosahedron respectively.

We can try to evaluate 
$\beta^{\rm eff}$ by using the perturbative expressions for the anomalous 
dimension of $I_n$ in the $O(3)$ nearest-neighbor lattice model. 
Explicit three-loop expressions are reported in 
App. \ref{app:anomale}. We obtain the following estimates
\begin{eqnarray}
\beta^{(3),\,{\rm eff}}  & = & 0.75, \label{3} \\
\beta^{(4),\,{\rm eff}}  & = & 1.08, \label{4} \\
\beta^{(6),\,{\rm eff}}  & = & 1.95,
\end{eqnarray}
which are quite stable with respect to the loop order. 
Thus, we expect that for $\beta < \beta^{(i),\,{\rm eff}}$ 
the breaking to the corresponding subgroup of $O(3)$
appears as irrelevant. 
Now, $\beta\approx 1$ corresponds to a very small correlation length.
Therefore, as soon as $\xi > 1$, one immediately realizes that the 
tetrahedron and the cubic model are not asymptotically free.
On the other hand, $\beta = 1.95$ corresponds to a quite large 
value of the correlation length $\xi = 166.7(4)$ \cite{Caracciolo:1995ud}.
Therefore, we expect the discrepancy to decrease steadily as $\beta$ 
increases, till values of $\xi$ of order\footnote{This number 
is simply an order of magnitude. Indeed, even if our estimates are quite 
stable with respect to $\beta$,
the correlation length varies rapidly with $\beta$. 
For instance, for $\beta = 1.90$ (resp.  2.00), which are very close
to our estimate of $\beta^{(6),\rm eff}$ we have
$\xi = 122.3(3)$ (228.5(7) resp.) \cite{Caracciolo:1995ud}.} 
200 and increase steadily 
afterwards. Thus, a clear signal of the difference between the two models
requires simulations on quite large lattices with $\xi \gg 200$. 

It is important to notice that these estimates are valid for perturbations 
of the lattice nearest-neighbor $O(3)$ action only. Indeed, 
as it should be expected, $\beta^{(n),\rm eff}$ is nonuniversal, 
being the solution of $\gamma^{(n)}(t^{\rm eff}) = 2$, where $\gamma^{(n)}(t)$
is the action-dependent anomalous dimension of $I_n$. This means that,
by changing the action, it is well possible that the system shows a 
non-$O(3)$-invariant behavior for much smaller values of $\xi$.

The argument given here applies quantitatively only for $h_n$ small.
Nonetheless, for the discrete models considered in the simulations
it represents a plausible scenario which 
explains the numerical results and is compatible with the 
standard theoretical framework used in the analyses of the 
critical behavior of the $O(3)$ model. 
Indeed, since all simulations are performed in the region in 
which $1 < \xi \ltapprox 100$, it predicts that the cubic and 
the tetrahedron model are clearly different from the $O(3)$ model,
while the icosahedron results should mimic the $O(3)$ ones.

%% Note also that the numerical results for the tetrahedron model
%% are very close to those for the octohedron model
%% \cite{Hasenfratz1,Hasenfratz2}. This is in agreement\footnote{
%% This statement is not a rigorous one. Indeed, we have proved 
%% that octahedral and tetrahedral perturbation are exactly equivalent 
%% at the $O(3)$ fixed point. Here instead, we are comparing the 
%% tetrahedron and the octahedron fixed points.} 
%% with our results which indicate that the important quantity is 
%% not the density of points on the sphere, but the dimension of the smallest
%% representation of the $O(3)$ group which is compatible with the reduced
%% symmetry.
  
It is interesting to notice that the values of $\beta^{(n),\rm eff}$ 
are close to the critical value for each discrete model 
$\beta^{(n)}_c$. For instance,
since the tetrahedron model is equivalent
to the 4-state Potts model with $\beta = 3/4\, \beta_{\hbox{\tiny
    Potts}}$ we have  
\be
\beta_c^{(3)} = {3\over 4} \log 3 \approx 0.82,
\ee
which is only slightly higher than \reff{3}.
Analogously, the cubic model is equivalent to the product of three Ising
models with $\beta = 3 \beta_{\hbox{\tiny Ising}}$ so that
\be
\beta_c^{(4)} = {3\over 2} \log (1+\sqrt{2})  \approx 1.32.
\ee
There is also~\cite{Patrascioiu:1991nn} a  numerical estimate for
$\beta_c^{(6)} \approx 2.15$.
Note also that $\beta^{(n),\,{\rm eff}} < \beta_c^{(n)}$, 
an inequality which shows that our $\beta^{(n),\,{\rm eff}}$ correspond always 
to temperatures above the freezing transition.

\section*{Acknowledgments}
A.M. thanks E.~Br\'ezin for a stimulating conversation on the subject of
this paper. We also thank A.~D.~Sokal for a careful reading of our manuscript,
P. Hasenfratz and F. Niedermayer for many useful comments.

\appendix

\section{Discrete subgroups of $O(3)$}

\subsection{The Platonic solids} \label{App1.1}

The icosahedron and the dodecahedron are two {\em Platonic
solids}. They are regular convex polyhedra~\cite{Coxeter}, with 
regular and equal faces, and are such that each vertex belongs to the 
same number of edges.
If its faces are $p$-gons (polygons with $p$ sides),
$q$ of them  surrounding each
vertex, the polyhedron is denoted by $\{p,q\}$. The possible
values for $p$ and $q$ may be enumerated as follows. The solid
angle at a vertex has $q$ face-angles, each $(p - 2) \pi/p$. Of
course, the sum of these $q$ angles must be less than $2\pi$. 
Therefore, we have
\be
{1\over q} + {1\over p} > {1\over 2}. \label{Plato}
\ee
Thus, $\{p,q\}$ cannot have any other values than
$$ \{3,3\}, \{3,4\}, \{4,3\}, \{3,5\}, \{5,3\}. $$
They correspond to the tetrahedron, the octahedron, the cube, the icosahedron,
and the dodecahedron.

Consider the regular polyhedron $\{p,q\}$ with its $N_0$ vertices,
$N_1$ edges, and $N_2$ faces, where  $N_0 -N_1 + N_2 =2$ by  Euler's
formula. As each face touches $p$  edges and each edge belongs to 2 faces, then
\be
p N_2 = 2 N_1.
\ee
Analogously, since
each vertex belongs to $q$ edges and each edge touches 2 vertices, we obtain
\be
q N_0 = 2 N_1 .
\ee 
These relations and Euler's formula imply
\be
{1\over q} + {1\over p} -{1\over 2} = {1\over N_1},
\ee
which offers a quantitative evaluation to the inequality~\reff{Plato}.

Consider now the sphere which touches all the edges. If we replace
each edge by a perpendicular line touching the sphere at the same
point, we obtain the $N_1$ edges of the {\em dual } polyhedron
$\{q,p\}$ which has $N_2$ vertices and $N_0$ faces.

We are interested in the rotation groups of the regular polyhedra.
They are finite groups, so that every rotation must have
an angle commensurable with $\pi$. In fact, the smallest angle of rotations
around a given axis is a submultiple of $2\pi$, and all other angles of
rotation about the same axis are multiples of the smallest one. If
$2\pi/k$ is the smallest angle, then the rotations about this axis
form a cyclic group of order $k$, and one speaks of an {\em axis of
$k$-fold rotation}.

Two dual  polyhedra have the same rotation group. The center of
the polyhedron 
$\{p,q\}$ is joined to the vertices, mid-edge points, and centers
of faces, by axes of $q$-fold, 2-fold, and $p$-fold rotation. But
the vertices, mid-edge points, and centers of faces occurs in
antipodal pairs. Hence, the total number of rotations, excluding
the identity, is 
$$ {1\over 2} \left[(q-1) N_0 + N_1 + (p-1) N_2
\right] = 2 N_1 -1, $$ 
so that the order of the rotation group is
$2N_1$.

The rotation group of the icosahedron  (and therefore of its dual
polyhedron, the dodecahedron) has 60 elements.

\subsection{The invariants under discrete subgroups}\label{inv}

In this Section we will compute the lowest-degree homogeneous polynomials 
in the $(x,y,z)$ coordinates that are invariant under the 
action of several discrete subgroups of $O(3)$. 
We will consider only the subgroups which are really
three-dimensional isometries: they are related to the Platonic solids
discussed above.
The other subgroups of proper rotations are the cyclic groups $C_n$,
for $n>1$,  and the dihedric 
groups $D_n$, for $n>2$, which are also subgroups of SO(2).

All invariants of $O(3)$ can be obtained as powers of the basic 
degree-two invariant
\be
I_2 = x^2 + y^2 + z^2 .
\ee
  
\subsubsection{The cubic group}

Let us fix the cube with vertices at the points $(\pm 1,\pm 1,\pm
1)/\sqrt{3}$. There are 48 matrices which leave invariant the cube,
that are
$$
\left( \begin{array}{rrr}\pm 1 & 0 & 0 \\ 0 & \pm 1 & 0 \\ 0 & 0 & \pm
      1 \end{array} \right), \quad
\left( \begin{array}{rrr} 0 & 0 & \pm
      1 \\\pm 1 & 0 & 0 \\ 0 & \pm 1 & 0  \end{array} \right), \quad
\left( \begin{array}{rrr} 0 & \pm 1 & 0  \\ 0 & 0 & \pm
      1 \\\pm 1 & 0 & 0  \end{array} \right),
$$
$$
\left( \begin{array}{rrr}\pm 1 & 0 & 0 \\ 0 & 0 & \pm
      1 \\ 0 & \pm 1 & 0 \end{array} \right) , \quad
\left( \begin{array}{rrr}  0 & \pm
      1 & 0 \\ \pm 1 & 0 & 0 \\ 0 & 0 &\pm 1   \end{array} \right) , \quad
\left( \begin{array}{rrr}0 & 0 & \pm
      1 \\0 & \pm 1  & 0 \\  \pm 1 & 0 & 0    \end{array} \right) .
$$
Twenty-four matrices ($2 N_1 = 24$) are proper rotations, while
the other 24 matrices are
obtained by compositions of proper rotations with the antipodal
transformation $\hbox{diag}(-1,-1,-1)$. Algebraically, the cubic
group is $S_4 \otimes \Z_2$, where $S_4$  is the group of permutations of 4
elements.
As the cube is dual to the octahedron the two groups of
invariance are the same.

It is easy to see that the lowest-order non-trivial polynomial is
\be
I_4 = x^4 + y^4 + z^4 + a\, I_2^2.
\ee
On the unit sphere the polynomials  can be decomposed  into
irreducible representations of O(3),  
i.e.  in terms of spherical harmonics. Then
\be
I_4 =  \left({3\over 5} + a \right) \sqrt{4 \pi} Y_{0,0} + {2\over 15}
\sqrt{4 \pi} 
Y_{4,0} + {2\over 3} \sqrt{4 \pi\over   70} \left( Y_{4,4} + Y_{4,-4} \right).
\ee
With the choice $a=-3/5$, we obtain an operator that is renormalized
multiplicatively.

\subsubsection{The group of the tetrahedron}

Let us choose the tetrahedron with vertices at the points
$$ 
{1\over \sqrt{3}} (1,1,1),\,{1\over \sqrt{3}} (1,-1,-1),\,{1\over
  \sqrt{3}} (-1,1,-1),\,{1\over \sqrt{3}} (-1,-1,1)\; .
$$
The matrices which leave invariant the tetrahedron are the 24 matrices 
of the cubic
group which have an even number of $-1$. Remark that there is not the
antipodal transformation, but there are 12 proper rotations and 12
reflections.
The lowest non-trivial polynomial is
\be
I_3 = x\,y\,z = - i\, \sqrt{ 2 \pi\over 105} \left( Y_{3,2} - Y_{3,-2}
\right) .
\ee
Algebraically this group is $A_4$, the group of even permutations of 4
elements.

\subsubsection{The group of the icosahedron}
We shall parametrize the 12 vertices of the icosahedron as
follows:
\begin{eqnarray}
P_u & = & (0,0,1), \\
P_d & = & (0,0,-1),\nonumber \\
P_k & = & \left( {2\over \sqrt{5} }\, \cos \left(  { \pi \, k\over
      5 }\right),  {2\over \sqrt{5} }\, \sin \left(  { \pi \, k\over
      5 }\right),  {1\over \sqrt{5} } \cos \left(  k\, \pi
  \right) \right), \nonumber 
\end{eqnarray}
for $k=1,\cdots,10$. 
In order to construct invariant polynomials under the rotation group of the 
icosahedron, we first consider
the cyclic group of order 5 of rotations of $2 \pi  / 5$ around the
$z$-axis. On the vertices it acts as a 
permutation of the form 
$$ (P_u)(P_1 P_3 P_5 P_7 P_9) (P_2 P_4 P_6 P_8 P_{10}) (P_d) .$$ 
In order to determine the invariants under
this cyclic group, it is convenient to use cylindric coordinates
$(z,\rho,\phi)\in R\times R^+ \times [0,2\pi ]$ so that the
action of the generator is
\be
(z,\rho,\phi) \to \left(z,\rho,\phi + {2 \pi \over 5}\right),
\ee 
and thus the invariants are $z$, $\rho$ and $5\phi$.
The lowest-order non-trivial polynomial which also respect the
invariance under the antipodal transformation  is of sixth degree and has
the general form 
\be
I_6 = z \left( z^5 + a z^3 \rho^2 + b z \rho^4 + c \rho^5 \cos (5
  \phi) + d \rho^5 \sin (5 \phi) \right) + e I_2^3,
\ee
where we have fixed to one the coefficient of $z^6$.
Of course, $I_6$ must be the same on all the vertices of the
icosahedron:
this gives the condition
\be
a = 31 - 4 b - 8 c.
\ee
We shall then use
the cyclic group of order 3 of rotations around the axis which
joins the origin with the center of the face $P_u P_2 P_4$. 
On the vertices it  acts as a permutation:
$$ (P_u P_2 P_4)(P_1 P_5 P_8)(P_3 P_6 P_{10})(P_7 P_9 P_d). $$ 
It gives the conditions
\begin{eqnarray}
b & = & 5, \\
c & = & 2, \\
d & = & 0,
\end{eqnarray}
so that $a = -5$. We obtain finally
\be
I_6 = z \left( z^5 -5  z^3 \rho^2 + 5 z \rho^4 + 2 \rho^5 \cos (5
  \phi)  \right) + e \left(z^2 + \rho^2\right)^3
\ee
or, in Cartesian coordinates,
\begin{eqnarray}
I_6 & = & z \left( z^5 -5  z^3 (x^2+y^2) + 5 z (x^2+y^2)^2 + 2 x (x^4 - 10
  x^2 y^2 + 5 y^4)  \right)\nonumber  \\
& & + e \left(x^2+y^2+z^2 \right)^3.
\end{eqnarray}
It can be checked that the all the other transformations, like the
cyclic groups of order 2 around the center of an edge, for example the
transformation which acts as a permutation like
$$ (P_u P_2)(P_1 P_6)(P_3 P_8)(P_4 P_{10})(P_5 P_9)(P_7 P_d), $$
leave $I_6$ invariant. Also the antipodal transformation leaves  $I_6$
invariant. 

On the unit sphere, in terms of spherical coordinates, we obtain
\be
I_6 = \left( {5\over 21} + e \right) \sqrt{4 \pi} Y_{0,0} + {16\over 21}
\sqrt{4 \pi\over 13} Y_{6,0} - {16\over 
  3} \sqrt{4 \pi\over 1001} \left( Y_{6,5} - Y_{6,-5}\right )   .
\ee
Therefore, for
\be
e = -  {5\over 21} \label{choice}
\ee
we obtain an operator which is multiplicatively renormalized.

Of course we could use a different position of the icosahedron in
space, for example we could take for the vertices
\be
{1\over \sqrt{2+\tau}}\,(0,\pm \tau,\pm 1)\quad 
{1\over \sqrt{2+\tau}}\,( \pm \tau,\pm 1, 0)\quad {1\over \sqrt{2+\tau}}\,
(\pm 1, 0, \pm \tau)
\ee
where $\tau$ is {\em golden ratio}. In this basis the cyclic group of
rotations of order five is generated by
\be
{1\over 2}\, \left( \begin{array}{ccc} \tau & \tau-1 & -1 \\  \tau-1 & 1  &
    \tau \\ 1 & -\tau & \tau -  1 \end{array} \right).
\ee
The new choice can be recovered from the old one by a rotation and
this  produces a different polynomial which turns out to be 
\begin{eqnarray}
I_6^\prime & = & \left( {5\over 21} + e \right) \sqrt{4 \pi}\, Y_{0,0} -
5\, \sqrt{4 \pi\over 273} \left[ {1\over \sqrt{21}}\, Y_{6,0} - {1\over
      2}  
\left( Y_{6,2} + Y_{6,-2}\right)\right. \nonumber \\
&& \left.  - {1\over \sqrt{6}} \left( Y_{6,4} +
  Y_{6,-4}\right) + {1\over 2} \sqrt{5\over 11}  \left( Y_{6,6} + 
    Y_{6,-6}\right)
 \right] , 
\end{eqnarray}
which, of course, with the choice \reff{choice} belongs to the same
multiplet $l=6$.

The icosahedron is dual to the dodecahedron and thus their group of
invariance is the same.

\section{The perturbative results}
\label{app:anomale}

In~\cite{Caracciolo:1994sk} the anomalous dimension of all non-derivative 
dimension-zero operators was computed for the nearest-neighbor 
lattice $O(N)$ $\sigma$-model up to
three loops. For generic $N$,
a suitable basis for these operators is given by
\be
{\cal O}^{(n)}_{j_1\ldots j_n}\, =\, \bsigma^{j_1}\ldots
\bsigma^{j_n}\, -\, \hbox{\rm traces} ,
\ee 
where ``traces" must be
such that ${\cal O}^{(n)}_{j_1\ldots j_n}$ is completely symmetric
and traceless. These polynomials are irreducible $O(N)$-tensors of
rank $n$ and as such they renormalize multiplicatively with no
off-diagonal mixing. For $N=3$ this representation is equivalent to that 
of the spherical harmonics.

The anomalous dimension $\gamma^{(n)}(t)$ of these operators is given by
\be
\gamma^{(n)}(t)\, =\, {\gamma^{(n)}_0 t}\, +\,
{\gamma^{(n)}_1 t^2}\, +\,
{\gamma^{(n)}_2 t^3}\, +\, O(t^4) ,
\ee 
where
\begin{eqnarray}
\gamma^{(n)}_0  &=& {n(N+n-2)\over4\pi},\nonumber \\ 
\gamma^{(n)}_1 &=& {n(N+n-2)\over16\pi},\\ 
\gamma^{(n)}_2 &=&
{n(N+n-2)\over4\pi}\left[ {N-2\over4\pi^2} \left(4 \pi^2 G_1 +
{1\over2} - {\pi^2\over8} \right)\, +\, {11\over96}  \right], \nonumber
\end{eqnarray}
where
\be
G_1  =  -{1\over4}\int \limits_{[-\pi,\pi]^d} {{\rm d}^d p \over
(2\pi)^d} {{\rm d}^d q \over (2\pi)^d} \left[\sum_\mu
\widehat{(p+q)^4_\mu}\right]\, {\widehat{(p+q)^2} - \hat{p}^2 -
\hat{q}^2 \over \hat{p}^2 \hat{q}^2 [\widehat{(p+q)^2}]^2 },
 \ee
and 
\be \hatp^2  \;=\; \sum_\mu \hatp^2_\mu  \;=\;
               \sum_\mu \left( 2 \sin {p_\mu\over 2} \right)^2    \;.
\ee
Numerically $G_1 \approx 0.0461636$\ . 
For $N=3$  we obtain 
\begin{eqnarray}
\gamma^{(3)}(t) &\, = \, &{0.95493t} + {0.238732t^2}
  + {0.135755t^3},  \\
\gamma^{(4)}(t) &\, = \, &{1.59155t} + {0.397887t^2}
  + {0.226258 t^3},  \\
\gamma^{(6)}(t) &\, = \, &{3.34225t} + {0.835563t^2}
  + {0.475142 t^3} .
\end{eqnarray}

\eject

\end{document}